\documentclass[preprint,12pt]{elsarticle}
\def\rf#1{(\ref{#1})}

\usepackage{amssymb}

\journal{Physics Letters B}

\begin{document}

\begin{frontmatter}



\title{Generalising the coupling between spacetime and matter.}


\author{Sante Carloni}

\address{Centro Multidisciplinar de Astrofisica - CENTRA,
Instituto Superior Tecnico - IST,
Universidade de Lisboa - UL,
Avenida Rovisco Pais 1, 1049-001, Portugal}

\begin{abstract}
We explore the idea that the coupling between matter and spacetime is more complex than the one originally envisioned by Einstein. We propose that such coupling takes the form of a new fundamental tensor in the Einstein field equations. We then show that the introduction of this tensor can account for dark phenomenology in General Relativity, maintaining a weak field limit compatible with standard Newtonian gravitation. The same paradigm can be applied any other theory of gravitation. We show, as an example, that in the context of conformal gravity a generalised coupling is able to solve compatibility issues between the matter and the gravitational sector. 
\end{abstract}


\end{frontmatter}

\section{Introduction}
One of the most important foundational assumptions of General Relativity (GR) is related with the way in which the curvature of spacetime and the matter in that spacetime interact. In his effort to construct a consistent theory, Einstein \cite{Einstein} chose a proportionality relation between what would be called the Einstein tensor and  the stress energy tensor of matter, intended as a continuum. This assumption was suggested by the requirement to have Poisson like equations for the gravitational field, the variational structure of the vacuum equations, but most of all requiring that the newly proposed gravitational field equations would contain naturally the conservation equations for the matter energy continuum.

Almost a hundred years later, the evidence on the evolution of the Universe at cosmological and astrophysical  scale has lead the research community to hypothesise the existence of dark (i.e. not interacting electromagnetically) fluids  which are characterised by exotic thermodynamics. It is interesting that this dark phenomenology always arises in relation to the gravitational behaviour of collection of many particles (the matter continuum). This is in contrast with the fact that most accurate tests of GR (e.g. the light deflection tests, the perihelion shift of planetary orbits, the Nordtvedt effect \cite{Will} and even the recently discovered gravity waves signals \cite{Abbott:2016blz}) are related to the behaviour of a small number of (test) particles. In other words, so far we have only tested the vacuum Einstein equations. In fact, data on the gravitation of many particle systems always leaves space for an interpretation of data which entails some form of modification of the gravitational interaction, like in the case of inflation \cite{ModInflation}. It is difficult to refrain from noting here, in spite of the profound conceptual differences, the resemblance with quantum mechanical systems, in which collections of particles have a global behaviour different for the one of single particle. 

It is then possible that, in the same line of thinking, the key to understanding dark phenomenology is related to ``global modes'' of gravitation of collection of particles? And how could this idea be realised concretely in the framework of relativistic theories of gravitation?

In this letter we propose a possible way to implement such a model. The central idea is that the coupling between matter and spacetime (i.e. the relation between the Einstein tensor and the stress energy tensor) is more complex than a simple law of proportionality and it is described by an additional fundamental tensor. When this idea is applied to pure GR, the generalised coupling is able to generate the non-trivial thermodynamics of the dark fluids without introducing exotic fields. In particular, via reconstruction techniques, we will discover that the new theory solves naturally problems like cosmic acceleration and the flattening of rotation curves of the galaxies, while leaving the GR vacuum phenomenology untouched and retaining the standard Newtonian limit (modulo a rescaling of the gravitational constant). In the case of more complex gravitational theories the use of generalised coupling has other advantages. We will show briefly that, in the case of conformal gravity, the generalised coupling helps resolve issues of compatibility of this theory with non-conformally invariant sources.

Throughout this paper  we will use natural units ($c=8\pi G=1$) and Latin indices run from 0 to 3. The conventions are consistent with Wald \cite{Wald}. The symmetrisation over the indexes of a tensor is defined as $T_{(a b)}= 1/2\left(T_{a b}+T_{b a}\right)$.

\section{Modified General Relativity.}
Let us suppose that the interaction between the curvature of spacetime and the stress energy tensor of matter is described by a given tensor field $\chi_{ab}{}^{cd}$ so that the Einstein equations can be written as
\begin{equation}\label{M-equation}
G_{ab}=\chi_{ab}{}^{cd} T_{cd}\;,
\end{equation}
where $G_{ab}$ is the Einstein tensor and $T_{cd}$ the matter stress energy tensor. In \rf{M-equation} spacetime couples in a different way with the different thermodynamic properties of matter  and we can imagine the components of   $\chi_{ab}{}^{cd}$ to represent such coupling. As for the original GR equations, we assume that for the stress energy tensor the usual conservation laws hold $\nabla^c T_{cd}=0$. Then the Bianchi identities $\nabla^a G_{ab}=0$ imply 
\begin{equation}\label{Bianchi}
\left(\nabla^b \chi_{ab}{}^{cd}\right) T_{cd}+\chi_{ab}{}^{cd}\nabla^b T_{cd}=0.
\end{equation}
which can be seen as a propagation equation for $\chi_{ab}{}^{cd}$.

What are the properties of $\chi_{ab}{}^{cd}$? The symmetries of $G_{ab}$  and $T_{cd}$ require 
\begin{equation}
\chi_{(ab)(cd)}=\chi_{abcd}.
\end{equation}
We will also suppose that $\chi$ is invertible. i.e. it has no zero eigenvalues and that for a set of free falling shear free vorticity free observers it is diagonal.  Since $\chi_{abcd}$ represents the coupling between matter and spacetime it is natural to assume that this tensor is different from zero only in presence of matter. The simplest way to implement this condition is to suppose that $\chi_{abcd}=\chi_{abcd}(\mu_0)$ and that $\chi_{abcd}(0)=0$. As a consequence, if we contract both sides of \rf{M-equation} by the inverse of $\chi_{abcd}$, obtaining 
\begin{equation}
\chi^{ab}{}_{cd} G_{ab}=T_{cd}
\end{equation}
the above equation is only valid if  $T_{ab}\neq0$. 

These choices guarantees that the Schwarzschild solution, as well as the other relevant vacuum solutions of General Relativity, is completely unaffected. This fact, in turn, implies that the totality of the classical tests for relativity is automatically satisfied.  Other local tests based, for example, on the geodesic deviation,  would be irrelevant for the detection of the generalised coupling. In fact, since for a given observer $\chi_{ab}{}^{cd}$ is determined and it contracts the total energy momentum tensor there is no way to reveal its action locally by a single inertial observer. 
 
Things change when we compare  the results of gravitational experiments made by different observers. Since $\chi_{ab}{}^{cd}$ is a tensor we can expect different observers to detect different strengths of the gravitational coupling. Such change is the key property one should look for to test \rf{M-equation} against observations. As well known, the only way to perform measurements of a tensorial quantity in a relativistic system is to construct scalar quantities associated to this tensor. In this letter,  focus will be given to the simple case of $T_{cd}=\mu u_c u_d$ where $u_a$ ($u_a u^a<0$) is the four velocity of a given observer. In this case, the R. H. S. of (1) can be written as
\begin{equation}\label{RHS1}
\chi_{ab}{}^{cd} u_c u_d \mu =\left(\chi_1 u_a u_b + \frac{1}{3}\chi_2 h_{ab}\right)\mu
\end{equation}
where $h_{ab}$  is the projection tensor on the surface orthogonal to ${u}_a$ and
\begin{eqnarray}\label{ChiDef}
&\chi_1=\chi_{ab}{}^{cd} u^a u^b u_c u_d\;,\\
&\chi_2=\chi_{ab}{}^{cd} h^{ab} u_c u_d.
\end{eqnarray}
are the two scalars, which encode the effect of the generalised coupling in this case.  We already know the transformation properties of $\mu$  from the structure of $T_{cd}$, thus the transformation properties  \rf{RHS1} and the change of the strength of the gravitational interaction under a Lorentz boost can be understood studying the transformation properties of $\chi_1$ and  $\chi_2$ . 

Upon the passage between two slow moving observers i.e. upon the transformation 
\begin{equation}
\tilde{u}_a=\gamma(u_a+v_a)\;, \quad \gamma= (1-|v|^2)^{-1/2} \;,\quad |v|\ll1\;,
\end{equation}
one obtains, for the projection tensor,
\begin{equation}
\tilde{h}_{ab}={h}_{ab}+ (\gamma^2-1)u_au_b\;,
\end{equation}
and therefore
\begin{eqnarray}\label{ChiTrans}
&\tilde{\chi}_1=\gamma^4\chi_1\approx\chi_1\;,\\
&\tilde{\chi}_2=\chi_2+\gamma^2(\gamma^2-1)\chi_1\approx\chi_2\;.
\end{eqnarray}
Hence if we were to perform an experiment able to measure the gravitational constant in two non relativistic different frames, the above results tell us that it would not easily reveal changes in the gravitational coupling. 

A more efficient testing ground for the generalised coupling  is to consider experiments in accelerated frames. Since the velocity of these observers changes in time, the presence of generalised coupling would be evident as a variation of the gravitational coupling with the velocity of the frame. We can sample this effect applying the modified Lorentz transformations to map inertial to non rotating accelerated observers \cite{Gourgoulhon}
\begin{eqnarray}
&\tilde{u}_a=\Gamma\left[\left(1+\dot{u}_b n^{b}\right)u_a+v_a\right]\;, \\
&\Gamma= \left[\left(1+\dot{u}_b n^{b}\right)-|v|^2\right]^{-1/2} \;,
\end{eqnarray}
where $n_b$ is the projection of the geodesic deviation of the two observers on the inertial observer rest frame. The transformation laws for the quantities \rf{ChiDef} become
\begin{eqnarray}\label{ChiTransAcc}
&\tilde{\chi}_1=\Gamma^4\left(1+\dot{u}_b n^{b}\right)^4\chi_1\;,\\
&\tilde{\chi}_2=\chi_2+\Gamma^2\left(1+\dot{u}_b n^{b}\right)^2\left[\Gamma^2\left(1+\dot{u}_b n^{b}\right)^2-1\right]\chi_1\;,
\end{eqnarray}
which now depends on time. In the case of constant acceleration  these quantities increase when the velocity increases. Such effect implies that the coupling with gravity will become stronger and stronger as the velocity of the laboratory increases. 

The \rf{M-equation} implies that there exist only one frame in which the gravitational field equations look like the standard Einstein equations, the frame in which $\chi_{ab}{}^{cd}=\delta_{(a}{}^{(c}\delta_{b)}{}^{d)}$. We can think of this ``Einstein frame'' as a special frame for this reason, but in fact, at this level, it has no special characteristics. The main idea that we will examine here is that in the frame in which we make cosmological and astrophysical observation is {\it not}  the Einstein frame and therefore that $\chi_{ab}{}^{cd}$ has a more complex form than the one above. 

We now focus on the dynamical aspects of the generalised coupling. In other to understand the properties of dynamical gravitational systems one will have to solve the \rf{M-equation} together with the propagation equation \rf{Bianchi} which describes the evolution of $\chi_{abcd}$. At this stage we will not perform a complete analysis of the solutions of this system, but we will rather check if the new theory is able to offer a theoretical framework for dark phenomenology without generating deviation form the gravitational phenomenology at Newtonian level. In the following, we will accomplish this task using reconstruction techniques \cite{Reconstruction}.   In particular, we will show that  in cosmology and in Newtonian gravitation the theory is able to generate a dark energy era and to flatten the rotation curves of the galaxies (intended as Newtonian systems) at the price of a rescaling of the gravitational coupling  at small scales.

\paragraph{Dark Cosmology}
Let us consider the simple case of a homogeneous and isotropic cosmological spacetime. We choose, as usual, a frame characterised by the geodesic observers attached to the particles of the cosmic fluid: the fundamental observers.  As mentioned in the previous section for these observers (free falling, shear free, vorticity free), the tensor $\chi_{abcd}$ is diagonal and  the metric is given by the well known Friedmann-Lema\^{\i}tre-Robertson-Walker (FLRW) line element.

Assuming that the cosmic fluid is made essentially of pressureless matter (i.e. $T_{cd}=\mu u_c u_d$)  ($u_a$ is now the four velocity of the fundamental observers) and calling $S$ the scale factor and $k$ the spatial curvature, the gravitational field equations and the Bianchi identities reduce to 
\begin{eqnarray}\label{Cosmo Eq}
&3 H^2 +3\frac{k}{S^2}=\chi_1 \mu\;,\quad \dot{H}-H^2=-\frac{1}{6}(\chi_1- 3\chi_2) \mu\;,\\
&\dot{\chi }_1+H\left(3\chi_1+\chi_2\right) =0\;,\quad\dot{\mu}+3H\mu=0\;,
\end{eqnarray}
where $H= \dot{S}/s$ and $\chi_1,\chi_2$ are given in \rf{ChiDef}. Note that $\chi_2$ is the result of the non trivial coupling between spacetime and matter and reduces to zero in the Einstein frame. It also evident that the system is not closed and, not differently form the pressure of a perfect fluid in standard cosmology,  the time dependence of $\chi_2$ has to be ascribed to some other relation. We can imagine this relation as an internal structural law of the spacetime itself, analogous to the equation of state of a perfect fluid.  

Let us attempt to reconstruct the form of the coupling able to give a relevant cosmic history.  From the cosmological equations \rf{Cosmo Eq}, one obtains
\begin{equation}
\chi_1=\frac{3 S }{\mu _0}\left(\dot{S}^2+k\right)\;,\quad \chi_2= -\frac{6 S}{\mu _0} \left(S \ddot{S}+2 \dot{S}^2+2 k\right)\;,
\end{equation}
where we have substituted the energy density with the solution of the continuity equation $\mu=\mu_0 S^{-3}$.

Starobinski and Sahni \cite{Sahni:1999gb} have proposed a solution of the cosmological equations that describes a spatially flat cosmology that evolves from a dust dominated solution to a cosmological constant dominated one. For this solution, the scale factor is $a=a_0 \sinh^{2/3} (\alpha t)$  and we have 
\begin{eqnarray}
&\chi_1=\frac{4 \alpha ^2 }{3 \mu _0}\cosh ^2(\alpha  t) \;,\\ 
&\chi_2=-\frac{4}{\mu _0} \left[3 k \sinh ^{\frac{2}{3}}(\alpha  t)+\alpha ^2 \cosh (2 \alpha 
   t)\right]\;.
\end{eqnarray}
Naturally the composition of the cosmic fluid is much more complex than simple dust and therefore this estimation should be extended to more complex fluids. Although more complicated algebraically, such extension does not present any conceptual problem.

\paragraph{Newtonian gravitation}
Let us now calculate the Newtonian limit of the field equations \rf{M-equation}. We will suppose a weak gravitational field  so that the metric $g_{ab}$ can always be written as $g_{ab}=\eta_{ab}+f_{ab}$ where $||f||\ll1$. We also choose  a set of  quasi-Newtonian observers with four velocity $\bar{u}_a$ which  are defined as shear free, vorticity free and accelerated $(\dot{\bar{u}}_b \neq0$) \cite{vanElst:1998kb}. We also assume these observers have very small three velocities and acceleration.

Following the standard procedure we write \rf{M-equation} as
\begin{equation}
R_{ab}=\chi_{ab}{}^{cd} T_{cd}-\frac{1}{2}g_{ab}\left(g^{ef}\chi_{ef}{}^{cd} T_{cd}\right).
\end{equation}
Considering that for any Newtonian distribution of matter $p\ll \mu$  and that in the Newtonian limit only the projection of both the free indexes of  this equation on the four velocity is relevant, we can write
\begin{equation}\label{NewtEq}
R_{ab}\bar{u}^a \bar{u}^b=\frac{\mu}{2}\left(\bar{\chi}_1+\bar{\chi}_2\right)\;,
\end{equation}
where we have used the definitions \rf{ChiDef}. The quantity on the left hand side of the previous equation in our conventions can be approximated to
\begin{equation}
R_{ab}\bar{u}^a \bar{u}^b\approx-\eta^{\mu\nu} \Phi_{;\mu\nu}\;,
\end{equation}
where $ \Phi=\frac{1}{2}f_{ab}\bar{u}^a \bar{u}^b$. Then the \rf{NewtEq} becomes the modified Poisson equation
\begin{equation}\label{Poisson}
 \Delta \Phi \approx\left(\chi_1+\chi_2\right)\mu\;,
\end{equation}
where we used the fact that for $|v|\ll1 ,|\dot{u}|\ll1 $ the \rf{ChiTransAcc} gives $\bar{\chi}_1\approx \chi_1$ and $\bar{\chi}_2\approx \chi_2$.

From this equation it appears clear that if the spacetime variation of the coefficients $\chi_i$ on cosmic/astrophysics scale is small compared to the time/distance scales of the Newtonian processes, they can be considered constants. In other words, the theory above on typical small scale Newtonian system just renormalises the value of the gravitational constant. 
\paragraph{Galaxy rotation curves.}
What happens to bigger system, e.g. above the galaxy scale? The formal solution of the \rf{Poisson} in a given set of Cartesian coordinates is 
\begin{eqnarray}
\Phi=\int \frac{\left(\chi_1+\chi_2\right)\mu}{|\mathbf{x}-\mathbf{x}_0|}d^{3}x\;,
\end{eqnarray}
where now $\chi_1$ and $\chi_2$ can be functions of the spatial coordinates. As in the case of cosmology, one can reconstruct the form of $\chi_1$ and $\chi_2$ which gives rise to the characteristic phenomenology associated to dark phenomenology e.g. the flattening of the rotation curves. For example, setting
\begin{equation}
\rho=\left(\frac{r}{r_0}\right)^{-\gamma}\;, \qquad \chi_1+\chi_2= \left[1+\left(\frac{r}{r_0}\right)^\alpha\right]^{\frac{\gamma-\beta}{\alpha}}\;,
\end{equation}
where $r$ is some radial coordinate, an isothermal halo can gravitate as a double power law density distribution  \cite{MvdBW}. Note that this form of $\chi_1$ and $\chi_2$ goes to a constant for $r_0\rightarrow\infty$ i.e. for small scales, thus the structure of the Poisson equation  at Solar System scales is still the \rf{Poisson}. 
  
It is important to stress that an interpretation of the rotation curves like the one above implies that the potential profile will be universal for all galaxies. It is known that in fact differences in those profiles can arise \cite{MvdBW}. However, since these galaxies contain matter with different equations of state, the result above should be properly extended in order to analyse in full this problem.

\section{Discussion and conclusions.}
In this letter we give a first exploration of the idea that the coupling between spacetime and matter is more complex than the one originally proposed by Einstein. This idea can be implemented by considering  a modified version of the Einstein equations in which a new field $\chi_{ab}{}^{cd}$ appears, which described the gravitational interaction. The dynamics of $\chi_{ab}{}^{cd}$  is given by an equation derived by the  Bianchi identities and complements the Einstein equations.  

The system of these equations can be quite complicated and to perform a fist exploration of the properties of the theory we made use of reconstruction techniques in the simple case of a pressureless source fluid. In this way we have been able to show that this extension of GR allows to give full control  over the behaviour of the matter interactions at different scales by introducing appropriate dimensional constants. Such degree of control can be used to offer an alternative theoretical framework for a number of unexplained phenomena like the realisation of cosmic acceleration and the flattening of the rotation curves of galaxies.  In the highly symmetric cases we have considered above it seems to emerge that there is a certain degree of degeneration in the different components of $\chi_{ab}{}^{cd}$. Such degeneration might indicate that some further simplifying assumption of the form of this tensor can be made. The understanding of the nature of these assumptions, however, requires a more detailed analysis of the theory, which is left for a future work. 

On the other hand, observing the effects of generalised coupling  locally is not an easy task. Measurements between inertial observers,  are not very useful to show evidence of non trivial coupling between spacetime and matter. By means of generalised Lorentz transformations, however, we showed that gravitational experiments in relativistic accelerated frames can be useful to accomplish this goal. 

The additional freedom of a generalised matter coupling can be exploited in different ways in different contexts. For example, one can solve immediately  the problem of the coupling of conformal gravity \cite{Mannheim:2011ds} with non conformal matter sources.  In fact, supposing that $\mathcal{S}$ is a conformally invariant functional which gives rise, upon variation, to  $\chi_{ab}{}^{cd} T_{cd}$  the conformal invariance will require $g^{ab}\chi_{ab}{}^{cd} T_{cd}=0$ not $g^{cd} T_{cd}=0$. Therefore conformal invariance does not imply necessarily a trace free stress energy tensor and the modified field equations for conformal gravity are consistent with any type of matter source. In this sense generalised coupling can be used as an alternative mechanism (for other strategies see \cite{Mannheim2006}) to couple conformal gravity to matter. In addition, it is not difficult to prove that the modification of the field equation does not affect any crucial  phenomenological property of this theory, such as the emergence of a dark energy era and the flattening of galaxy rotation curves.

How are theories with generalised coupling different from other proposals of modification/extension of GR? The \rf{M-equation}, for example, is a purely classical  theory in the sense that no quantum information is encoded in its structure.  It also does not entail any other degree of freedom of spacetime (e.g. torsional ones), although a generalisation of this type is straightforward. Mathematically the \rf{M-equation} might resemble a theory in which matter is non minimally coupled to gravitation, like the Brans-Dicke theory \cite{BD} or the more recently proposals in \cite{BT} or \cite{Carroll:2006jn} and one can make use of the \rf{M-equation} to study these models (and indeed any other modification of GR) in a generalised framework. In this sense, one can say that the \rf{M-equation}  contains multiple modifications of GR.  However it should be stressed that there is an important physical difference between these theories and the theory proposed in this work. The tensor $\chi_{abcd}$ in \rf{M-equation} is not conceived here as a special matter field. There is no stress energy tensor associated to $\chi_{abcd}$ in \rf{M-equation} and, most of all, $\chi_{abcd}$ is related to the matter content of the theory in such a way that the vacuum sector of the new theory it is exactly GR. 

Since the introduction of a generalised coupling implies the presence of new parameters in the theory, it should come to no surprise that such an extensions can give account of dark phenomenology.  This might be considered a shortcoming of this idea. It should be remarked however that the totality of proposals for interpretation of dark phenomenology implies the introduction of some additional degrees of freedom, thus \rf{M-equation} is not ``worse'' than any other proposal for modification of GR. The crucial advantage of \rf{M-equation} with respect to those models is the testability of the underlying idea of generalised coupling and the agreement with all the test of Einsteinian gravitation presently available. Further studies will shed more light on the physics of generalised coupling allowing an even more detailed constraint of this idea against observations.

\vspace{0.3cm}
{\bf Acknowledgments:} This work was supported by  the Funda\c{c}\~{a}o para a Ci\^{e}ncia e Tecnologia through project IF/00250/2013.  I also acknowledge financial support provided under the European Union's H2020 ERC Consolidator Grant ``Matter and strong-field gravity: New frontiers in Einstein's theory'' grant agreement no.  MaGRaTh?646597, and under the H2020-MSCA-RISE-2015 Grant No. StronGrHEP-690904.





\end{document}